\documentclass[twocolumn,times]{aastex6}
\usepackage{amsmath}
\usepackage[nolist]{acronym}

\providecommand{\svnid}[1]{} 

\newcommand{\macro}[1]{\textcolor{red}{#1}}

\newcommand{\Msun}{\ensuremath{\mathrm{M}_\odot}}

\newcommand\OBSEVENTDATE{\macro{2015 September 14}}
\newcommand\OBSEVENTTIME{\macro{09:50:45}}

     \newcommand{\MONESCOMPACT}{\macro{\ensuremath{36_{-4}^{+5}}}}  \newcommand{\MTWOSCOMPACT}{\macro{\ensuremath{29_{-4}^{+4}}}}  \newcommand{\MFINALSCOMPACT}{\macro{\ensuremath{62_{-4}^{+4}}}}      \newcommand{\DISTANCECOMPACT}{\macro{\ensuremath{410_{-180}^{+160}}}} \newcommand{\REDSHIFTCOMPACT}{\macro{\ensuremath{0.09_{-0.04}^{+0.03}}}}

\newcommand{\FIXME}[1]{\textcolor{red}{#1}}
\renewcommand{\FIXME}[1]{#1}
\renewcommand{\macro}[1]{#1}

\begin{document}

\title{Localization and broadband follow-up of the gravitational-wave transient GW150914}

\slugcomment{The Astrophysical Journal Letters, 826:L13, 2016 July 20}
\received{2016 February 29}
\accepted{2016 April 26}
\published{2016 July 20}

\defcitealias{GCN18330}{18330}
\defcitealias{GCN18331}{18331}
\defcitealias{GCN18332}{18332}
\defcitealias{GCN18333}{18333}
\defcitealias{GCN18334}{18334}
\defcitealias{GCN18335}{18335}
\defcitealias{GCN18336}{18336}
\defcitealias{GCN18337}{18337}
\defcitealias{GCN18338}{18338}
\defcitealias{GCN18339}{18339}
\defcitealias{GCN18340}{18340}
\defcitealias{GCN18341}{18341}
\defcitealias{GCN18343}{18343}
\defcitealias{GCN18344}{18344}
\defcitealias{GCN18345}{18345}
\defcitealias{GCN18346}{18346}
\defcitealias{GCN18347}{18347}
\defcitealias{GCN18348}{18348}
\defcitealias{GCN18349}{18349}
\defcitealias{GCN18350}{18350}
\defcitealias{GCN18353}{18353}
\defcitealias{GCN18354}{18354}
\defcitealias{GCN18359}{18359}
\defcitealias{GCN18361}{18361}
\defcitealias{GCN18362}{18362}
\defcitealias{GCN18363}{18363}
\defcitealias{GCN18364}{18364}
\defcitealias{GCN18370}{18370}
\defcitealias{GCN18371}{18371}
\defcitealias{GCN18372}{18372}
\defcitealias{GCN18388}{18388}
\defcitealias{GCN18390}{18390}
\defcitealias{GCN18394}{18394}
\defcitealias{GCN18395}{18395}
\defcitealias{GCN18397}{18397}
\defcitealias{GCN18420}{18420}
\defcitealias{GCN18424}{18424}
\defcitealias{GCN18474}{18474}
\defcitealias{GCN18655}{18655}
\defcitealias{GCN18690}{18690}
\defcitealias{GCN18709}{18709}
\defcitealias{GCN18851}{18851}
\defcitealias{GCN18858}{18858}
\defcitealias{GCN18903}{18903}
\defcitealias{GCN18914}{18914}
\defcitealias{GCN19013}{19013}
\defcitealias{GCN19017}{19017}
\defcitealias{GCN19021}{19021}
\defcitealias{GCN19022}{19022}
\defcitealias{GCN19034}{19034}
 \newcommand\citegcn[1]{\citetalias{GCN#1}}

\AuthorCallLimit=-1
\author{The~LIGO~Scientific~Collaboration~and~the~Virgo~Collaboration}
\author{the~Australian~Square~Kilometer~Array~Pathfinder~(ASKAP)~Collaboration}
\author{the~BOOTES~Collaboration}
\author{the~Dark~Energy~Survey~and~the~Dark~Energy~Camera~GW-EM~Collaborations}
\author{the~\emph{Fermi}~GBM~Collaboration}
\author{the~\emph{Fermi}~LAT~Collaboration}
\author{the~GRAvitational~Wave~Inaf~TeAm~(GRAWITA)}
\author{the~\emph{INTEGRAL}~Collaboration}
\author{the~Intermediate~Palomar~Transient~Factory~(iPTF)~Collaboration}
\author{the~InterPlanetary~Network}
\author{the~J-GEM~Collaboration}
\author{the~La~Silla--QUEST~Survey}
\author{the~Liverpool~Telescope~Collaboration}
\author{the~Low~Frequency~Array~(LOFAR)~Collaboration}
\author{the~MASTER~Collaboration}
\author{the~MAXI~Collaboration}
\author{the~Murchison~Wide-field~Array~(MWA)~Collaboration}
\author{the~Pan-STARRS~Collaboration}
\author{the~PESSTO~Collaboration}
\author{the~Pi~of~the~Sky~Collaboration}
\author{the~SkyMapper~Collaboration}
\author{the~\emph{Swift}~Collaboration}
\author{the~TAROT,~Zadko,~Algerian~National~Observatory,~and~C2PU~Collaboration}
\author{the~TOROS~Collaboration}
\author{the~VISTA~Collaboration}
\affil{See the Supplement, \citealt{GW150914-EMFOLLOW-SUPPLEMENT}, for the full list of authors.\\}
\email{lsc-spokesperson@ligo.org}
\email{virgo-spokesperson@ego-gw.eu}

\shorttitle{Localization and broadband follow-up of GW150914}
\shortauthors{Abbott et al.}


\begin{abstract}
A gravitational-wave (\acsu{GW}) transient was identified in data recorded by the Advanced Laser Interferometer Gravitational-wave Observatory (\acsu{LIGO}) detectors on {\OBSEVENTDATE}.
The event, initially designated G184098 and later given the name GW150914, is described in detail elsewhere.
By prior arrangement, preliminary estimates of the time, significance, and sky location of the event were shared with \FIXME{63} teams of observers covering radio, optical, near-infrared, X-ray, and gamma-ray wavelengths with ground- and space-based facilities.
In this Letter we describe the low-latency analysis of the \ac{GW} data and present the sky localization of the first observed compact binary merger.
We summarize the follow-up observations reported by \FIXME{25} teams via private \acl{GCN} circulars, giving an overview of the participating facilities, the \ac{GW} sky localization coverage, the timeline and depth of the observations.
As this event turned out to be a \acl{BBH} merger, there is little expectation of a detectable \ac{EM} signature.
Nevertheless, this first broadband campaign to search for a counterpart of an Advanced \ac{LIGO} source represents a milestone and highlights the broad capabilities of the transient astronomy community and the observing strategies that have been developed to pursue \acl{NS} binary merger events. Detailed investigations of the \ac{EM} data and results of the \ac{EM} follow-up campaign are being disseminated in papers by the individual teams.
\end{abstract}

\acresetall

\acused{BAYESTAR}

\section{Introduction}
\label{s:introduction}

In 2015 September, the Advanced Laser Interferometer Gravitational-wave Observatory (\acsu{LIGO}; \citealt{aLIGO}) made the first direct detection of an astrophysical gravitational-wave (\acsu{GW}) signal that turned out to be from a \ac{BBH} merger.
The LIGO Hanford and Livingston sites are the first two nodes of a growing global network of highly sensitive \ac{GW} facilities, soon to include Advanced Virgo \citep{AdVirgo}, KAGRA, and LIGO\nobreakdashes--India.
Some of the most promising astrophysical sources of \acp{GW} are also expected to produce broadband \ac{EM} emission and neutrinos.
This has created exciting new opportunities for joint broadband \ac{EM} observations and multi-messenger astronomy.

In a \ac{CBC} event, a tight binary comprised of two \acp{NS}, two \acp{BH}, or a \ac{NS} and a \ac{BH} experiences a runaway orbital decay due to gravitational radiation.
In a binary including at least one \ac{NS}---
a \ac{BNS} or \ac{NSBH} merger---we expect \ac{EM} signatures due to energetic outflows at different timescales and wavelengths.
If a relativistic jet forms, we may observe a prompt short \ac{GRB} lasting on the order of one second or less, followed by X\nobreakdashes-ray, optical, and radio afterglows of hours to days duration (e.g., \citealt{Eichler:1989,Narayan:1992,Nakar:2007,Berger:2014,Fong:2015}).
Rapid neutron capture in the sub-relativistic ejecta (e.g., \citealt{Lattimer:1976}) is hypothesized to produce a \acl{KN} or macronova, an optical and near-infrared signal lasting hours to weeks (e.g., \citealt{LP98}).
Eventually, we may observe a radio blast wave from this sub\nobreakdashes-relativistic outflow, detectable for months to years (e.g., \citealt{np11}).
Furthermore, several seconds prior to or tens of minutes after merger, we may see a coherent radio burst lasting milliseconds (e.g., \citealt{Hansen:2001,Zhang:2014}).
In short, a \ac{NS} binary can produce \ac{EM} radiation over a wide range of wavelengths and time scales.
On the other hand, in the case of a stellar-mass \ac{BBH}, the current consensus is that no significant \ac{EM} counterpart emission is expected except for those in highly improbable environments pervaded by large ambient magnetic fields or baryon densities.

The first campaign to find \ac{EM} counterparts triggered by low-latency \ac{GW} event candidates was carried out with the initial LIGO and Virgo detectors and several \ac{EM} astronomy facilities in 2009 and 2010 \citep{S6lowlatencyCBC,S6EMmethods,S6Swift,S6opticalEM}. 
In preparing for Advanced detector operations, the LIGO and Virgo collaborations worked with the broader astronomy community to set up an evolved and greatly expanded \ac{EM} follow-up program.\footnote{See program description and participation information at \url{http://www.ligo.org/scientists/GWEMalerts.php}.}
\FIXME{Seventy-four} groups with access to ground- and space-based facilities joined, of which \FIXME{63} were operational during \ac{O1}.
Details of the 2009 to 2010 \ac{EM} follow campaign and changes for O1 are given in Section~1 of the Supplement \citep{GW150914-EMFOLLOW-SUPPLEMENT}.

After years of construction and commissioning, the Advanced LIGO detectors at Livingston, Louisiana, and Hanford, Washington, began observing in 2015 September with about $3.5$ times the distance reach ($>40$ times the sensitive volume) of the earlier detectors.
A strong \ac{GW} event was identified shortly after the pre-run calibration process was completed. Deep analysis of this event, initially called G184098 and later given the name GW150914, is presented in \citet{GW150914-DETECTION} and companion papers referenced therein.
In this paper we describe the initial low-latency analysis and event candidate selection (Section~\ref{s:detection}), the rapid determination of likely sky localization (Section~\ref{s:skymaps}), and the follow-up \ac{EM} observations carried out by partner facilities (Sections~\ref{s:emfollow}~and~\ref{s:limits}).
For analyses of those observations, we refer the reader to the now-public \ac{GCN}~circulars\footnote{All circulars related to GW150914 are collected at \url{http://gcn.gsfc.nasa.gov/other/GW150914.gcn3}}
and to a number of recent papers.
We end with a brief discussion of \ac{EM} counterpart detection prospects for future events.

\section{Data Analysis and Discovery}
\label{s:detection}

As configured for \ac{O1}, four low-latency pipelines continually search for transient signals that are coincident in the two detectors within the $10$\,ms light travel time separating them.
\acl{cWB} (\acsu{cWB}; \citealt{cWB-overview}) and \acl{oLIB} (\acsu{oLIB}; \citealt{oLIB}) both search for unmodeled \ac{GW} bursts \citep{GW150914-BURST}. GSTLAL \citep{LLOID,GW150914-GSTLAL}
and \acl{MBTA} (\acsu{MBTA};~\citealt{MBTA}) search specifically for \ac{NS} binary mergers using matched filtering.
Because \ac{CBC} waveforms can be precisely computed from general relativity, GSTLAL and \ac{MBTA} are more sensitive to \ac{CBC} signals than the burst search pipelines are.
All four detection pipelines report candidates within a few minutes of data acquisition.

LIGO conducted a series of engineering runs throughout Advanced LIGO's construction and commissioning to prepare to collect and analyze data in a stable configuration.
The \ac{ER8} began on 2015~August~17 at 15:00 and critical software was frozen by August 30.\footnote{All dates and times are in UT.}
The rest of \ac{ER8} was to be used to calibrate the detectors, to carry out diagnostic studies, to practice maintaining a high coincident duty cycle, and to train and tune the data analysis pipelines.
Calibration was complete by September~12 and \ac{O1} was scheduled to begin on September~18.
On {\OBSEVENTDATE}, \ac{cWB} reported a burst candidate to have occurred at {\OBSEVENTTIME} with a network \ac{SNR} of 23.45 and an estimated \ac{FAR} $< 0.371$ yr$^{-1}$ based on the available (limited at that time) data statistics.
Also, \ac{oLIB} reported a candidate with consistent timing and \ac{SNR}.
No candidates were reported at this time by the low-latency GSTLAL and \ac{MBTA} pipelines, ruling out a \ac{BNS} or \ac{NSBH} merger.

Although the candidate occurred before \ac{O1} officially began, the LIGO and Virgo collaborations decided to send an alert to partner facilites because the preliminary \ac{FAR} estimate satisfied our planned alert threshold of 1\,month$^{-1}$.
Although we had not planned to disseminate real-time \ac{GCN} notices before the formal start of \ac{O1}, most of the computing infrastructure was in place.
Basic data quality checks were done within hours of GW150914; both interferometers were stable and the data stream was free of artifacts \citep{GW150914-DETCHAR}.
A \ac{cWB} sky map was available 17 minutes after the data were recorded, and a \ac{LIB} sky map was available after 14\,hr.
After extra data integrity checks and an update to the \ac{GCN} server software, these two sky maps were communicated to observing partners in a \ac{GCN} circular nearly two days after the event occurred (GCN~\citegcn{18330}).
Mass estimates were not released in this initial circular, and observers may have assumed the event was associated with a \ac{BNS} system or a \ac{GW} burst (e.g., from a nearby core-collapse \acl{SN}; \acsu{SN}).
The knowledge that GW150914 was consistent with a \ac{BBH} inspiral and merger was only shared later on October~3 (GCN~\citegcn{18388}).
\begin{figure*}
    \includegraphics[width=\textwidth]{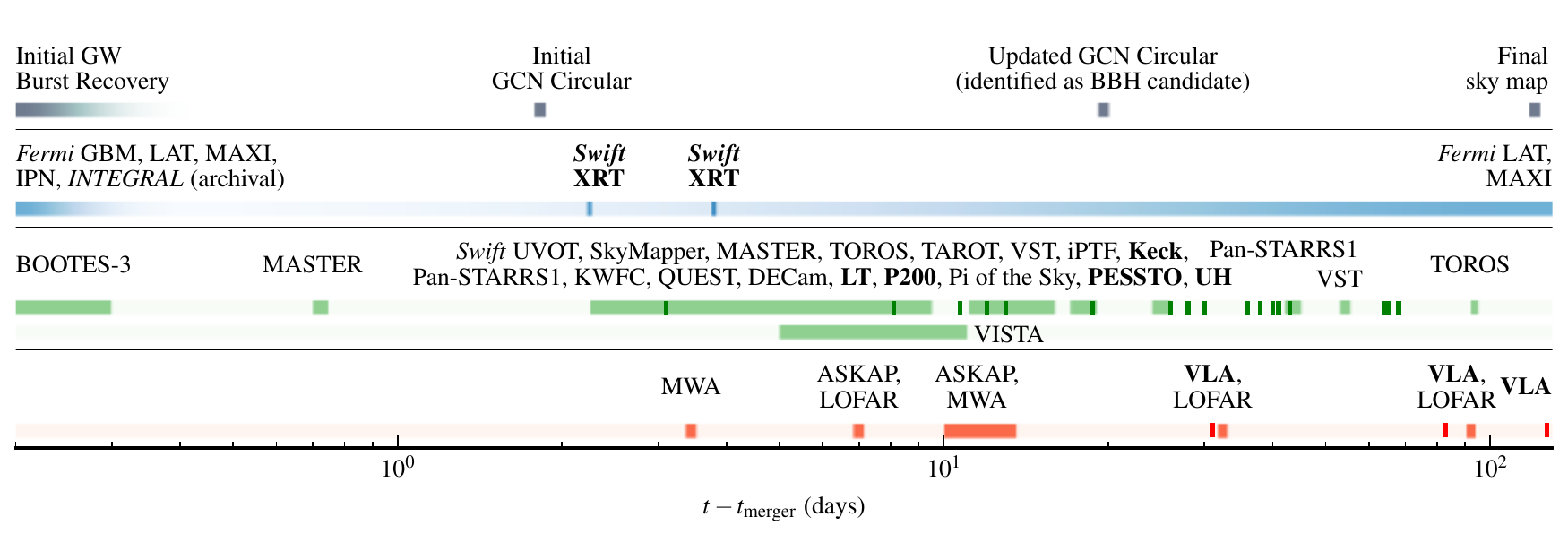}
    \caption{\label{fig:timeline}
        Timeline of observations of GW150914, separated by band and relative to the time of the \ac{GW} trigger.
        The top row shows \ac{GW} information releases.
        The bottom four rows show high-energy, optical, near-infrared, and radio observations, respectively. Optical spectroscopy and narrow-field radio observations are indicated with darker tick marks and boldface text.
        Table~\ref{table:footprints} reports more detailed information on the times of observations made with each instrument.
    }
\end{figure*}
Figure~\ref{fig:timeline} shows the chronology of the \ac{GW} detection alerts and follow-up observations.

The data were re-analyzed offline with two independent matched-filter searches using a template bank that includes both \ac{NS} binary and \ac{BBH} mergers.
The waveform was confirmed to be consistent with a \ac{BBH} merger and this information was shared with observers about three weeks after the event (GCN~\citegcn{18388}).
The \ac{FAR} was evaluated with the data collected through 20 October, reported to be less than 1 in 100 years (GCN~\citegcn{18851}; \citealt{GW150914-CBC}), and ultimately determined to be much lower.
The final results of the offline search are reported in \citet{GW150914-DETECTION}.

\section{Sky Maps}
\label{s:skymaps}

\begin{figure*}[t]
    \includegraphics[width=\textwidth]{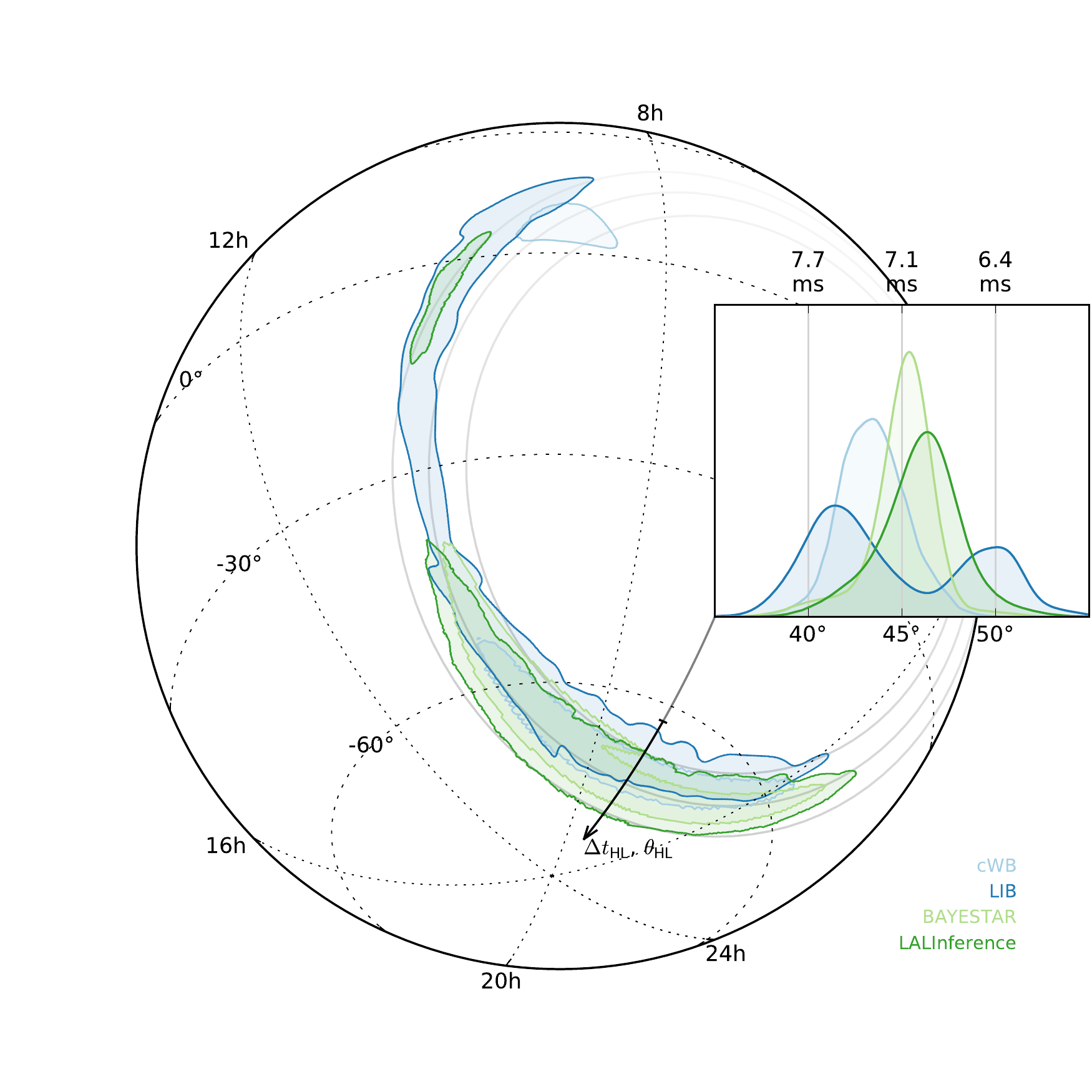}
    \caption{\label{fig:contours}
        Comparison of different \ac{GW} sky maps, showing the 90\% credible level contours for each algorithm.
        This is an orthographic projection centered on the centroid of the \ac{LIB} localization.
        The inset shows the distribution of the polar angle $\theta_\mathrm{HL}$ (equivalently, the arrival time difference $\Delta t_\mathrm{HL}$).
    }
\end{figure*}

We produce and disseminate probability sky maps using a sequence of algorithms with increasing accuracy and computational cost.
Here, we compare four location estimates: the prompt \ac{cWB} and \ac{LIB} localizations that were initially shared with observing partners plus the rapid \ac{BAYESTAR} localization and the final localization from LALInference.
All four are shown in Fig.~\ref{fig:contours}.

\ac{cWB} performs a constrained \acl{ML} estimate of the reconstructed signal on a sky grid \citep{cWB-overview} weighted by the detectors' antenna patterns \citep{BF2Y} and makes minimal assumptions about the waveform morphology.
With two detectors, this amounts to restricting the signal to only one of two orthogonal \ac{GW} polarizations throughout most of the sky. \ac{LIB} performs Bayesian inference assuming the signal is a sinusoidally modulated Gaussian \citep{oLIB}.
While this assumption may not perfectly match the data, it is flexible enough to produce reliable localizations for a wide variety of waveforms, including \ac{BBH} \acl{IMR} signals \citep{BF2Y}.
\acs{BAYESTAR} produces sky maps by triangulating the times, amplitudes, and phases on arrival supplied by all the \ac{CBC} pipelines \citep{BAYESTAR}. \acs{BAYESTAR} was not available promptly because the low-latency \ac{CBC} searches were not configured for \acp{BBH}; the localization presented here is derived from the offline \ac{CBC} search.
LALInference performs full forward modeling of the data using a parameterized \ac{CBC} waveform which allows for \ac{BH} spins and detector calibration uncertainties \citep{LALInference}.
It is the most accurate method for CBC signals but takes the most time due to the high dimensionality.
We present the same LALInference map as \citet{GW150914-PARAMESTIM}, with a spline interpolation procedure to include the potential effects of calibration uncertainties.
The \ac{BAYESTAR} and LALInference maps were shared with observers on 2016~January~13 (GCN~\citegcn{18858}), at the conclusion of the \ac{O1} run.
Since GW150914 is a \ac{CBC} event, we consider the LALInference map to be the most accurate, authoritative, and \textit{final} localization for this event.
This map has a 90\% credible region with area 630 deg$^2$.

All of the sky maps agree qualitatively, favoring a broad, long section of arc in the southern hemisphere and to a lesser extent a shorter section of nearly the same arc near the equator.
While the majority of \ac{LIB}'s probability is concentrated in the southern hemisphere, a non-trivial fraction of the 90\% confidence region extends into the northern hemisphere.
The LALInference sky map shows much less support in the northern hemisphere which is likely associated with the stronger constraints available with full \ac{CBC} waveforms.
The \ac{cWB} localization also supports an isolated hot spot near $\alpha \sim 9^\mathrm{h}, \delta \sim 5\arcdeg$,
where the detector responses make it possible to independently measure two polarization components.
In this region, cWB considers signals not constrained to have the elliptical polarization expected from a compact binary merger.

Quantitative comparisons of the four sky maps can be found in section 2 of the Supplement \citep{GW150914-EMFOLLOW-SUPPLEMENT}.
The main feature in all of the maps is an annulus with polar angle $\theta_\mathrm{HL}$ determined by the arrival time difference $\Delta t_\mathrm{HL}$ between the two detectors.
However, refinements are possible due to phase as well as amplitude consistency and the mildly directional antenna patterns of the LIGO detectors \citep{Kasliwal:2014,F2Y}.
In particular, the detectors' antenna patterns dominate the modulation around the ring for un-modeled reconstructions through a correlation with the inferred distance of the source \citep{BF2Y}.
As shown in Fig.~\ref{fig:contours}, the algorithms all infer polar angles that are consistent at the $1\sigma$ level.

\section{Follow-up Observations}
\label{s:emfollow}

\begin{figure*}[t!]
    \includegraphics[width=\textwidth]{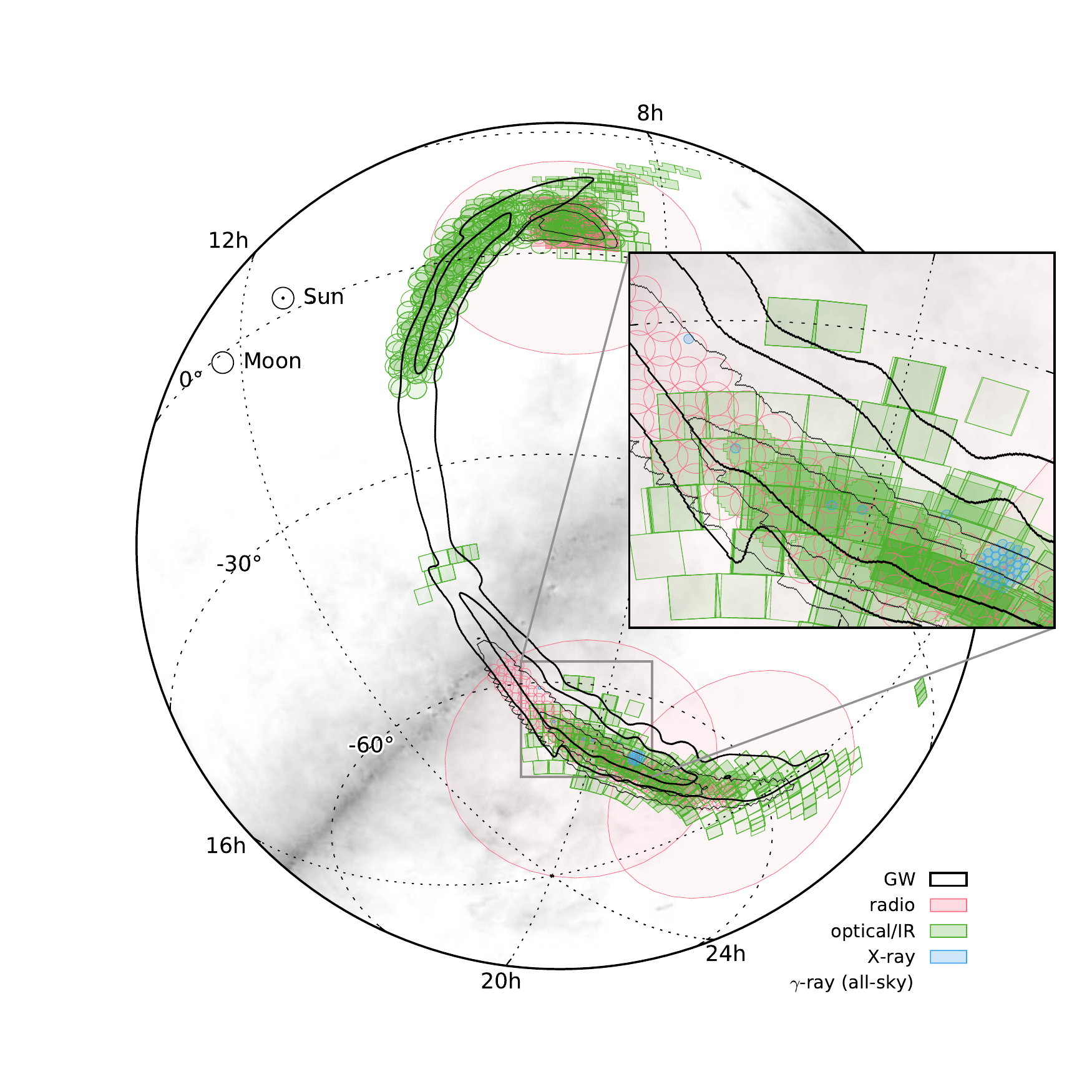}
    \caption{\label{fig:tiles}
        Footprints of observations in comparison with the 50\% and 90\% credible levels of the initially distributed \ac{GW} localization maps.
        Radio fields are shaded in red, optical/infrared fields are in green, and the \ac{XRT} fields are indicated by the blue circles.
        The all-sky \emph{Fermi} \ac{GBM}, \ac{LAT}, \emph{INTEGRAL} SPI\nobreakdashes-ACS, and MAXI observations are not shown.
        Where fields overlap, the shading is darker.
        The initial \ac{cWB} localization is shown as thin black contour lines and the \ac{LIB} localization as thick black lines.
        The inset highlights the \emph{Swift} observations consisting of a hexagonal grid and a selection of the \emph{a posteriori} most highly ranked galaxies.
        The \citet{SFDReddeningMap} reddening map is shown in the background to represent the Galactic plane.
        The projection is the same as in Fig.~\ref{fig:contours}.
    }
\end{figure*}

\FIXME{Twenty-five} participating teams of observers responded to the \ac{GW} alert to mobilize satellites and ground-based telescopes spanning 19 orders of magnitude in \ac{EM} wavelength.
Observations and archival analysis started shortly after the candidate was announced, two days after the event was recorded.
Most facilities followed tiling strategies based on the \ac{cWB} and \ac{LIB} sky maps.
Some groups, considering the possibility of a \ac{NS} merger or core-collapse \ac{SN}, selected fields based on the areal density of nearby galaxies or targeted the \ac{LMC} (e.g., \citealt{GW150914_DES_1}).
Had the \ac{BBH} nature of the signal been promptly available, most groups would not have favored local galaxies because \acs{LIGO}'s range for \ac{BBH} mergers is many times larger than that for \acp{BNS}.
Fig.~\ref{fig:tiles} displays the footprints of all reported observations. The campaign is summarized in Table~\ref{table:footprints} in terms of instruments, depth, time, and sky coverage. Some optical candidate counterparts were followed up spectroscopically and in the radio band as summarized in Table~\ref{table:followup}. The overall \ac{EM} follow-up  of GW150914 consisting of broad-band tiled observations and observations to characterize candidate counterparts are described in detail in Sections~3 through~5 of the Supplement \citep{GW150914-EMFOLLOW-SUPPLEMENT}.

Findings from these follow-up observations have been reported in several papers.
A weak transient signal was found in \emph{Fermi} GBM data 0.4\,s after the time of GW150914 \citep{GW150914_GBM}, but no corresponding signal was found in the \emph{INTEGRAL} SPI\nobreakdashes--ACS instrument \citep{GW150914_INTEGRAL} or \emph{AGILE} \citep{GW150914_AGILE}.
No GRB-like afterglow was found in X-rays with \emph{Swift} XRT \citep{GW150914_Swift} or MAXI (N.~Kawai~et~al., in preparation), in UV/optical with \emph{Swift} UVOT \citep{GW150914_Swift}, or at GeV energies with \emph{Fermi} LAT \citep{GW150914_LAT}.
Tiled observations with wide-field optical instruments listed in Table~\ref{table:footprints} found many transients, but spectroscopy with the instruments listed in Table~\ref{table:followup} along with further photometry showed that none of them were associated with GW150914 \citep{GW150914_iPTF,GW150914_pan_STARRPESSTO,GW150914_DES_2,GW150914_JGEM}.
\citet{GW150914_DES_1} used DECam to search for a missing supergiant in the \ac{LMC}, which would have been evidence for the collapse of a massive star that could have produced \acp{GW}, but failed to produce a typical core-collapse \ac{SN}.

\section{Coverage}
\label{s:limits}

Using the \ac{GW} data alone, we can only constrain the position of the source on the sky to an area of $\approx$600\,deg$^2$ (90\% confidence).  The inferred redshift is $z = \REDSHIFTCOMPACT$, corresponding to a luminosity distance of \DISTANCECOMPACT\,Mpc \citep{GW150914-PARAMESTIM}.

Table~\ref{table:footprints} lists the area tiled by each facility and the probability contained within those tiles, calculated with respect to the localization methods described in Section~\ref{s:skymaps}.

By far the most complete coverage of the area is at the highest energies.
The \emph{INTEGRAL} SPI\nobreakdashes--ACS provided the largest effective area in the 75\,keV\nobreakdashes--1\,MeV range, albeit with significantly varying detection efficiency. Owing to its nearly omnidirectional response, it had a full coverage of the GW probability map (GCN~\citegcn{18354}; \citealt{GW150914_INTEGRAL}). \emph{Fermi} \ac{GBM} captured 75\% of the localization at the time of the \ac{GW} trigger and the entire area by 25 minutes after (GCN~\citegcn{18339}). \emph{Fermi} LAT observations started 4200\,s after the trigger and the entire localization continued to be observed every three hours.

Coverage in X\nobreakdashes-rays is complete down to 10$^{-9}$\,erg\,cm$^{-2}$\,s$^{-1}$ with the MAXI observations, but
relatively sparse at fainter flux, with the \emph{Swift} \ac{XRT} tiles spanning about 5\,deg$^2$ and enclosing a probability of \FIXME{$\sim$0.3\%} in the energy range 0.3\nobreakdashes--10\,keV to a depth of 10$^{-13}$\nobreakdashes--10$^{-11}$\,erg\,cm$^{-2}$\,s$^{-1}$ (GCNs~\citegcn{18331}, \citegcn{18346}).

Optical facilities together tiled about \FIXME{900}\,deg$^2$ and captured a containment probability of over \FIXME{50\%} of the initial \ac{LIB} sky map and slightly less for the final LALInference sky map that was available after the observations were completed.
The depth varies widely among these facilities.
MASTER and Pan\nobreakdashes-STARRS1 covered the most area with their observations, while large areas also were covered by the \ac{IPTF}, \ac{DECAM}, \acl{VST} (\acsu{VST}; E. Brocato et al. 2016, in preparation), and La~Silla--QUEST.
The contained probability of the initial sky maps is dominated by MASTER, \ac{DECAM}, Pan\nobreakdashes-STARRS1, La~Silla--QUEST, and \ac{VST}, while the final sky map is contained best by MASTER, \ac{DECAM} and \ac{VST}.
Relatively small area and contained probability were covered by facilities that targeted nearby galaxies. The only wide-field near-infrared facility, the \ac{VISTA}, covered \FIXME{70}\,deg$^2$ and captured a containment probability of \FIXME{8\%} of the final LALInference sky map.

The radio coverage is also extensive, with a contained probability of \FIXME{86\%}, dominated by \ac{MWA} in the 118\,MHz band (GCN~\citegcn{18345}).

Table~\ref{table:followup} lists the observations done by large telescope spectrographs  and a radio facility to follow-up candidate optical counterparts.
Deep photometry, broadband observations, and spectroscopy identified the majority of the candidates to be normal population type Ia and type II \acp{SN}, with a few dwarf novae and \acp{AGN} that are all very likely unrelated to GW150914.
Candidate classification, comparison of redshift with the GW distance, and use of source age are crucial constraints to rule candidates in and out.
Detailed discussions of candidate selection, spectroscopic and broadband follow-up are presented in survey-specific publications about iPTF candidates \citep{GW150914_iPTF} and about PESSTO follow-up of Pan\nobreakdashes-STARRS1 candidates \citep{GW150914_pan_STARRPESSTO}.

\begin{deluxetable*}{lccc|r|rrrr|l}
\tablecaption{\label{table:footprints}Summary of Tiled Observations}
\tablewidth{\textwidth}
\tablecolumns{10}
\tablehead{
    \colhead{Facility/} &
    \colhead{} &
    \colhead{} &
    \colhead{} &
    \colhead{Area} &
    \multicolumn{4}{c}{Contained Probability (\%)} &
    \colhead{}
    \\
    \cline{6-9}
    \colhead{Instrument} &
    \colhead{Band\tablenotemark{a}} &
    Depth\tablenotemark{b} &
    \colhead{Time\tablenotemark{c}} &
    \colhead{(deg$^2$)} &
    \colhead{cWB} &
    \colhead{LIB} &
    \colhead{BSTR\tablenotemark{d}} &
    \colhead{LALInf} &
    \colhead{GCN}
    }
\startdata
\\[-1.5em]\cutinhead{Gamma-ray}
\emph{Fermi} LAT	&20\,MeV--	&$1.7\times10^{-9}$	&(every	&---	&100	&100	&100	&100	&\citegcn{18709}\\
	&300\,GeV	&	&3\,hr)	&	&	&	&	&	&\\
\emph{Fermi} GBM	&8\,keV\nobreakdashes--40\,MeV	&0.7--$5\times10^{-7}$	&(archival)	&---	&100	&100	&100	&100	&\citegcn{18339}\\
	&	&(0.1--1\,MeV)	&	&	&	&	&	&	&\\
INTEGRAL	&75\,keV--1\,MeV	&$1.3\times10^{-7}$	&(archival)	&---	&100	&100	&100	&100	&\citegcn{18354}\\
IPN	&15\,keV--10\,MeV	&$1\times10^{-7}$	&(archival)	&---	&100	&100	&100	&100	&---\\
\cutinhead{X-ray}
MAXI/GSC	&2--20\,keV	&$1\times10^{-9}$	&(archival)	&17900	&95	&89	&92	&84	&\citegcn{19013}\\
\emph{Swift} XRT	&0.3--10\,keV	&$5\times10^{-13}$ (gal.)	&2.3, 1, 1	&0.6	&0.03	&0.18	&0.04	&0.05	&\citegcn{18331}\\
	&	&2--$4\times10^{-12}$ (LMC)	&3.4, 1, 1	&4.1	&1.2	&1.9	&0.16	&0.26	&\citegcn{18346}\\
\cutinhead{Optical\tablenotemark{e}}
DECam	&$i, z$	&$i<22.5$, $z<21.5$	&3.9, 5, 22	&100	&38	&14	&14	&11	&\citegcn{18344}, \citegcn{18350}\\
iPTF	&$R$	&$R<20.4$	&3.1, 3, 1	&130	&2.8	&2.5	&0.0	&0.2	&\citegcn{18337}\\
KWFC	&$i$	&$i<18.8$	&3.4, 1, 1	&24	&0.0	&1.2	&0.0	&0.1	&\citegcn{18361}\\
MASTER	&C	&$<19.9$	&-1.1, 7, 7	&710	&50	&36	&55	&50	&\citegcn{18333}, \citegcn{18390}, \citegcn{18903}, \citegcn{19021}\\
Pan-STARRS1	&$i $	&$i < 19.2-20.8$	&3.2, 21, 42	&430	&28	&29	&2.0	&4.2	&\citegcn{18335}, \citegcn{18343}, \citegcn{18362}, \citegcn{18394}\\
La Silla--	&$g, r$ 	&$r < 21$	&3.8, 5, 0.1	&80	&23	&16	&6.2	&5.7	&\citegcn{18347}\\
\;\;QUEST	& 	&	&	&	&	&	&	&	&\\
SkyMapper	&$i, v$	&$i<19.1$, $v<17.1$	&2.4, 2, 3	&30	&9.1	&7.9	&1.5	&1.9	&\citegcn{18349}\\
\emph{Swift} UVOT	&$u$	&$u<19.8$ (gal.)	&2.3, 1, 1	&3	&0.7	&1.0	&0.1	&0.1	&\citegcn{18331}\\
	&$u$	&$u<18.8$ (LMC)	&3.4, 1, 1	&	&	&	&	&	&\citegcn{18346}\\
TAROT	&C	&$R<18$	&2.8, 5, 14	&30	&15	&3.5	&1.6	&1.9	&\citegcn{18332}, \citegcn{18348}\\
TOROS	&C	&$r<21$	&2.5, 7, 90	&0.6	&0.03	&0.0	&0.0	&0.0	&\citegcn{18338}\\
VST@ESO	&$r$	&$r<22.4$	&2.9, 6, 50	&90	&29	&10	&14	&10	&\citegcn{18336}, \citegcn{18397}\\
\cutinhead{Near Infrared}
VISTA@ESO	&$Y, J, K_S$	&$J<20.7$	&4.8, 1, 7	&70	&15	&6.4	&10	&8.0	&\citegcn{18353}\\
\cutinhead{Radio}
ASKAP	&863.5\,MHz	&5--15\,mJy	&7.5, 2, 6	&270	&82	&28	&44	&27	&\citegcn{18363}, \citegcn{18655}\\
LOFAR	&145\,MHz	&12.5\,mJy	&6.8, 3, 90	&100	&27	&1.3	&0.0	&0.1	&\citegcn{18364}, \citegcn{18424}, \citegcn{18690}\\
MWA	&118\,MHz	&200\,mJy	&3.5, 2, 8	&2800	&97	&72	&86	&86	&\citegcn{18345}\\
 \enddata
\tablenotetext{a}{Band: photon energy, optical or near-infrared filter (or C for clear unfiltered light), wavelength range, or central frequency.}
\tablenotetext{b}{Depth: gamma/X-ray limiting flux in erg\,cm$^{-2}$\,s$^{-1}$; $5\sigma$ optical/IR limiting magnitude (AB); and $5\sigma$ radio limiting spectral flux density in mJy. The reported values correspond to the faintest flux/magnitude of detectable sources in the images.}
\tablenotetext{c}{Elapsed time in days between start of observations and the time of GW150914 (2015~September~14~09:50:45), number of repeated observations of the same area, and total observation period in days.}
\tablenotetext{d}{BAYESTAR.}
\tablenotetext{e}{Searches for bright optical transients were also done by BOOTES-3 and Pi of the Sky. Details are given in the Supplement \citep{GW150914-EMFOLLOW-SUPPLEMENT}.}
\end{deluxetable*}

\begin{deluxetable*}{lclllll}[ht!]
\tablecaption{\label{table:followup}Summary of Follow-up Observations}
\tablewidth{\textwidth}
\tablecolumns{7}
\tablehead{\multicolumn{7}{c}{Spectroscopic Follow-up}}
\startdata
Instrument &  No. of Candidates & Discovery Survey & Epochs &  $\lambda$ (\AA) & $\Delta\lambda$\tablenotemark{a} (\AA)  & GCN\\
\tableline
KeckII+DEIMOS &  8 & iPTF & 1 & $4650-9600$        & $3.5$  & \citegcn{18337}, \citegcn{18341}\\
LT+SPRAT & 1 & Pan-STARRS1 & 1 & $4500-7500$ & $18$ & \citegcn{18370}, \citegcn{18371}\\
NTT+EFOSC2        &  10 & QUEST/Pan-STARRS1 & 1 & $3650-9250$ & $18$ & \citegcn{18359}, \citegcn{18395}\\
P200+DBSP          &  1 &  Pan-STARRS1 & 1& $3200-9000$ &  $4-8$ & \citegcn{18372}\\
UH2.2m+SNIFS         & 9 & Pan-STARRS1 & 1 & $3200-10000$ & $4-6$ & --- \\
\tableline
\multicolumn{7}{c}{Radio Follow-up}\\
\tableline
Instrument & No. of Candidates & Discovery Survey & Epochs & Freq. (GHz) & Lim. Flux\tablenotemark{b}  (uJy) & GCN\\
\tableline
VLA   & 1  & iPTF & 3 & 6 & $\lesssim 50$ & \citegcn{18420}, \citegcn{18474}, \citegcn{18914} \\
 \enddata
\tablenotetext{a}{Full width at half maximum resolution.}
\tablenotetext{b}{$5\sigma$, 2 GHz nominal bandwidth, $\approx 20$\,min on-source.}
\end{deluxetable*}

\section{Sensitivity}
\label{s:sensitivity}

The third column of Table~\ref{table:footprints} summarizes the depth of the follow-up program. We provide limiting flux, flux density, or magnitude for the different facilities. 
We emphasize that these limits only apply to the fraction of the sky contours that have been followed up. 
For example, the \ac{MWA} fields have an 86\% chance of containing the
source's sky location and provide no constraints on sky locations representing the remaining 14\%.

Because the follow-up program was primarily designed to search for counterparts to
\ac{BNS} and \ac{NSBH} mergers, it is interesting to note that the observational
campaign would have provided powerful constraints on such a system.
A \ac{BNS} coalescence could have been detected by LIGO during O1 at a distance
of $\sim70$ Mpc, averaged over sky position and
orientations \citep{O1-SENSITIVITY}.
A short \ac{GRB} afterglow similar to those that have been observed
by \emph{Swift} XRT would have been detectable at that distance.
Re-scaling the observed X-ray fluxes of short GRBs at 11\,hr after the event
would yield fluxes in the range
$2\times10^{-11}$ to $6\times 10^{-8}\,$erg\,cm$^{-2}$\,s$^{-1}$ \citep{Berger:2014}.
Kilonova emission from a \ac{BNS} merger at that distance would reach a peak apparent magnitude of 17--24 within a week or two after the merger (e.g., \citealt{2010MNRAS.406.2650M,BarnesKasenKilonova,Tanaka:2013,grossman2014MNRAS}).
This range overlaps with the depth reached in the optical and near-IR bands.
Finally, this BNS system might produce radio emission from tens of $\mu$Jy to tens of mJy (e.g., \citealt{hp:2015}) with different timescales spanning weeks to years. Tables~\ref{table:footprints}~and~\ref{table:followup} show that the radio observations from wide-field facilities were sensitive to mJy flux densities at low frequencies where fainter sources with longer timescales are expected, while the narrow-field VLA was sensitive to well localized radio transients down to   $\mu$Jy flux densities at frequencies above a few GHz.

\section{Conclusions}
\label{s:conclusion}

GW150914 is consistent with the inspiral
and merger of two \acp{BH} of masses $\MONESCOMPACT$ and
$\MTWOSCOMPACT\,\Msun$, respectively, resulting in the formation of a
final \ac{BH} of mass $\MFINALSCOMPACT\,\Msun$ \citep{GW150914-DETECTION}. In classical general relativity, a \emph{vacuum} \ac{BBH} merger does not produce any \ac{EM} or particle emission whatsoever.
Whereas supermassive \acp{BBH} in galactic centers may appear as dual \acp{AGN} or have other distinctive \ac{EM} signatures due to interactions with gas or magnetic fields,
stellar \ac{BBH} systems are not expected to possess detectable EM counterparts. 
The background gas densities and magnetic field strengths should therefore be typical of the interstellar medium, which are many orders of magnitude smaller than the environments of \ac{EM} bright supermassive \acp{BBH}. Although GW150914 is loud in \acp{GW} and expected to be absent in all \ac{EM} bands,
thorough follow-up observations were pursued 
to check for \ac{EM} emission. Future EM follow-ups of GW sources will shed light on the presence or absence of firm EM counterparts and astrophysical processes that may trigger EM emission from these systems.

The \ac{EM} campaign following GW150914 successfully demonstrates the capability
of the observing partners to cover large swaths of the sky localization area, to
identify candidates, and to activate larger telescopes for photometric and
spectroscopic characterization within a few days of an event.
We note that the information about the source's BBH nature and updated
sky maps were sent out 20 days and four months after the event,
respectively. This resulted in some instruments covering much less of
the probability region or to the required depth of GW150914 than they
may have planned for. We expect future
alerts to be issued within tens of minutes with more information
about the signal type and more rapid updates of
the maps. The follow-up efforts would have been sensitive to a wide range of
emission expected from BNS or NSBH mergers.
However, the widely variable
sensitivity reached across the sky localization area continues to be a
challenge for an \ac{EM} counterpart search.

The number of galaxies \citep[with luminosities $L \geqslant 0.1
  L^{\star}$;][]{Blanton:2003} within the comoving volume of $10^{-2}$ Gpc$^{3}$
corresponding to the 90$\%$ credible area of the LALInference sky map and within
the 90$\%$ confidence interval distance is $\sim10^{5}$.  Such a number makes it
impossible to identify the host galaxy in the absence of an \ac{EM} counterpart
detection. The presence of a third \ac{GW} detector such as Virgo would have
improved the sky localization of GW150914 to a few tens of square degrees both
for the unmodeled and \ac{CBC} searches. The future addition of more
\ac{GW} detectors to the global network \citep{ObservingScenarios} will
significantly improve the efficiency of searches for \ac{EM} counterparts.

In summary, we have described the \ac{EM} follow-up program carried out for the first \ac{GW} source detected by Advanced \ac{LIGO}.
Within two days of the initial tentative detection of
GW150914, a \ac{GCN} circular was sent to \ac{EM}
follow-up partners alerting them to the event
and providing them with initial sky maps. 
\FIXME{Twenty-five} \ac{EM} observing teams mobilized their resources, and over the
ensuing three months observations were performed with a diverse array
of facilities over a broad wavelength range (from radio to $\gamma$-ray).
Findings from those observations are being disseminated in other papers.
The localization and broadband follow-up of this \ac{GW} event constitutes an important first step in a new era of gravitational wave multi-messenger astronomy.

\providecommand{\acrolowercase}[1]{\lowercase{#1}}

\begin{acronym}
\acro{2D}[2D]{two\nobreakdashes-dimensional}
\acro{2+1D}[2+1D]{2+1\nobreakdashes--dimensional}
\acro{3D}[3D]{three\nobreakdashes-dimensional}
\acro{2MASS}[2MASS]{Two Micron All Sky Survey}
\acro{AdVirgo}[AdVirgo]{Advanced Virgo}
\acro{AMI}[AMI]{Arcminute Microkelvin Imager}
\acro{AGN}[AGN]{active galactic nucleus}
\acroplural{AGN}[AGN\acrolowercase{s}]{active galactic nuclei}
\acro{aLIGO}[aLIGO]{Advanced \acs{LIGO}}
\acro{ASKAP}[ASKAP]{Australian \acl{SKA} Pathfinder}
\acro{ATCA}[ATCA]{Australia Telescope Compact Array}
\acro{ATLAS}[ATLAS]{Asteroid Terrestrial-impact Last Alert System}
\acro{BAT}[BAT]{Burst Alert Telescope\acroextra{ (instrument on \emph{Swift})}}
\acro{BATSE}[BATSE]{Burst and Transient Source Experiment\acroextra{ (instrument on \acs{CGRO})}}
\acro{BAYESTAR}[BAYESTAR]{BAYESian TriAngulation and Rapid localization}
\acro{BBH}[BBH]{binary black hole}
\acro{BHBH}[BHBH]{\acl{BH}\nobreakdashes--\acl{BH}}
\acro{BH}[BH]{black hole}
\acro{BNS}[BNS]{binary neutron star}
\acro{CARMA}[CARMA]{Combined Array for Research in Millimeter\nobreakdashes-wave Astronomy}
\acro{CASA}[CASA]{Common Astronomy Software Applications}
\acro{CFH12k}[CFH12k]{Canada--France--Hawaii $12\,288 \times 8\,192$ pixel CCD mosaic\acroextra{ (instrument formerly on the Canada--France--Hawaii Telescope, now on the \ac{P48})}}
\acro{CRTS}[CRTS]{Catalina Real-time Transient Survey}
\acro{CTIO}[CTIO]{Cerro Tololo Inter-American Observatory}
\acro{CBC}[CBC]{compact binary coalescence}
\acro{CCD}[CCD]{charge coupled device}
\acro{CDF}[CDF]{cumulative distribution function}
\acro{CGRO}[CGRO]{Compton Gamma Ray Observatory}
\acro{CMB}[CMB]{cosmic microwave background}
\acro{CRLB}[CRLB]{Cram\'{e}r\nobreakdashes--Rao lower bound}
\acro{cWB}[\acrolowercase{c}WB]{Coherent WaveBurst}
\acro{DASWG}[DASWG]{Data Analysis Software Working Group}
\acro{DBSP}[DBSP]{Double Spectrograph\acroextra{ (instrument on \acs{P200})}}
\acro{DCT}[DCT]{Discovery Channel Telescope}
\acro{DECAM}[DECam]{Dark Energy Camera\acroextra{ (instrument on the Blanco 4\nobreakdashes-m telescope at \acs{CTIO})}}
\acro{DES}[DES]{Dark Energy Survey}
\acro{DFT}[DFT]{discrete Fourier transform}
\acro{EM}[EM]{electromagnetic}
\acro{ER8}[ER8]{eighth engineering run}
\acro{FD}[FD]{frequency domain}
\acro{FAR}[FAR]{false alarm rate}
\acro{FFT}[FFT]{fast Fourier transform}
\acro{FIR}[FIR]{finite impulse response}
\acro{FITS}[FITS]{Flexible Image Transport System}
\acro{FLOPS}[FLOPS]{floating point operations per second}
\acro{FOV}[FOV]{field of view}
\acroplural{FOV}[FOV\acrolowercase{s}]{fields of view}
\acro{FTN}[FTN]{Faulkes Telescope North}
\acro{FWHM}[FWHM]{full width at half-maximum}
\acro{GBM}[GBM]{Gamma-ray Burst Monitor\acroextra{ (instrument on \emph{Fermi})}}
\acro{GCN}[GCN]{Gamma-ray Coordinates Network}
\acro{GMOS}[GMOS]{Gemini Multi-object Spectrograph\acroextra{ (instrument on the Gemini telescopes)}}
\acro{GRB}[GRB]{gamma-ray burst}
\acro{GSC}[GSC]{Gas Slit Camera}
\acro{GSL}[GSL]{GNU Scientific Library}
\acro{GTC}[GTC]{Gran Telescopio Canarias}
\acro{GW}[GW]{gravitational wave}
\acro{HAWC}[HAWC]{High\nobreakdashes-Altitude Water \v{C}erenkov Gamma\nobreakdashes-Ray Observatory}
\acro{HCT}[HCT]{Himalayan Chandra Telescope}
\acro{HEALPix}[HEALP\acrolowercase{ix}]{Hierarchical Equal Area isoLatitude Pixelization}
\acro{HEASARC}[HEASARC]{High Energy Astrophysics Science Archive Research Center}
\acro{HETE}[HETE]{High Energy Transient Explorer}
\acro{HFOSC}[HFOSC]{Himalaya Faint Object Spectrograph and Camera\acroextra{ (instrument on \acs{HCT})}}
\acro{HMXB}[HMXB]{high\nobreakdashes-mass X\nobreakdashes-ray binary}
\acroplural{HMXB}[HMXB\acrolowercase{s}]{high\nobreakdashes-mass X\nobreakdashes-ray binaries}
\acro{HSC}[HSC]{Hyper Suprime\nobreakdashes-Cam\acroextra{ (instrument on the 8.2\nobreakdashes-m Subaru telescope)}}
\acro{IACT}[IACT]{imaging atmospheric \v{C}erenkov telescope}
\acro{IIR}[IIR]{infinite impulse response}
\acro{IMACS}[IMACS]{Inamori-Magellan Areal Camera \& Spectrograph\acroextra{ (instrument on the Magellan Baade telescope)}}
\acro{IMR}[IMR]{inspiral-merger-ringdown}
\acro{IPAC}[IPAC]{Infrared Processing and Analysis Center}
\acro{IPN}[IPN]{InterPlanetary Network}
\acro{IPTF}[\acrolowercase{i}PTF]{intermediate \acl{PTF}}
\acro{ISM}[ISM]{interstellar medium}
\acro{ISS}[ISS]{International Space Station}
\acro{KAGRA}[KAGRA]{KAmioka GRAvitational\nobreakdashes-wave observatory}
\acro{KDE}[KDE]{kernel density estimator}
\acro{KN}[KN]{kilonova}
\acroplural{KN}[KNe]{kilonovae}
\acro{LAT}[LAT]{Large Area Telescope}
\acro{LCOGT}[LCOGT]{Las Cumbres Observatory Global Telescope}
\acro{LHO}[LHO]{\ac{LIGO} Hanford Observatory}
\acro{LIB}[LIB]{LALInference Burst}
\acro{LIGO}[LIGO]{Laser Interferometer \acs{GW} Observatory}
\acro{llGRB}[\acrolowercase{ll}GRB]{low\nobreakdashes-luminosity \ac{GRB}}
\acro{LLOID}[LLOID]{Low Latency Online Inspiral Detection}
\acro{LLO}[LLO]{\ac{LIGO} Livingston Observatory}
\acro{LMI}[LMI]{Large Monolithic Imager\acroextra{ (instrument on \ac{DCT})}}
\acro{LOFAR}[LOFAR]{Low Frequency Array}
\acro{LOS}[LOS]{line of sight}
\acroplural{LOS}[LOSs]{lines of sight}
\acro{LMC}[LMC]{Large Magellanic Cloud}
\acro{LSB}[LSB]{long, soft burst}
\acro{LSC}[LSC]{\acs{LIGO} Scientific Collaboration}
\acro{LSO}[LSO]{last stable orbit}
\acro{LSST}[LSST]{Large Synoptic Survey Telescope}
\acro{LT}[LT]{Liverpool Telescope}
\acro{LTI}[LTI]{linear time invariant}
\acro{MAP}[MAP]{maximum a posteriori}
\acro{MBTA}[MBTA]{Multi-Band Template Analysis}
\acro{MCMC}[MCMC]{Markov chain Monte Carlo}
\acro{MLE}[MLE]{\ac{ML} estimator}
\acro{ML}[ML]{maximum likelihood}
\acro{MOU}[MOU]{memorandum of understanding}
\acroplural{MOU}[MOUs]{memoranda of understanding}
\acro{MWA}[MWA]{Murchison Widefield Array}
\acro{NED}[NED]{NASA/IPAC Extragalactic Database}
\acro{NSBH}[NSBH]{neutron star\nobreakdashes--black hole}
\acro{NSBH}[NSBH]{\acl{NS}\nobreakdashes--\acl{BH}}
\acro{NSF}[NSF]{National Science Foundation}
\acro{NSNS}[NSNS]{\acl{NS}\nobreakdashes--\acl{NS}}
\acro{NS}[NS]{neutron star}
\acro{O1}[O1]{\acl{aLIGO}'s first observing run}
\acro{oLIB}[\acrolowercase{o}LIB]{Omicron+\acl{LIB}}
\acro{OT}[OT]{optical transient}
\acro{P48}[P48]{Palomar 48~inch Oschin telescope}
\acro{P60}[P60]{robotic Palomar 60~inch telescope}
\acro{P200}[P200]{Palomar 200~inch Hale telescope}
\acro{PC}[PC]{photon counting}
\acro{PESSTO}[PESSTO]{Public ESO Spectroscopic Survey of Transient Objects}
\acro{PSD}[PSD]{power spectral density}
\acro{PSF}[PSF]{point-spread function}
\acro{PS1}[PS1]{Pan\nobreakdashes-STARRS~1}
\acro{PTF}[PTF]{Palomar Transient Factory}
\acro{QUEST}[QUEST]{Quasar Equatorial Survey Team}
\acro{RAPTOR}[RAPTOR]{Rapid Telescopes for Optical Response}
\acro{REU}[REU]{Research Experiences for Undergraduates}
\acro{RMS}[RMS]{root mean square}
\acro{ROTSE}[ROTSE]{Robotic Optical Transient Search}
\acro{S5}[S5]{\ac{LIGO}'s fifth science run}
\acro{S6}[S6]{\ac{LIGO}'s sixth science run}
\acro{SAA}[SAA]{South Atlantic Anomaly}
\acro{SHB}[SHB]{short, hard burst}
\acro{SHGRB}[SHGRB]{short, hard \acl{GRB}}
\acro{SKA}[SKA]{Square Kilometer Array}
\acro{SMT}[SMT]{Slewing Mirror Telescope\acroextra{ (instrument on \acs{UFFO} Pathfinder)}}
\acro{SNR}[S/N]{signal\nobreakdashes-to\nobreakdashes-noise ratio}
\acro{SSC}[SSC]{synchrotron self\nobreakdashes-Compton}
\acro{SDSS}[SDSS]{Sloan Digital Sky Survey}
\acro{SED}[SED]{spectral energy distribution}
\acro{SGRB}[SGRB]{short \acl{GRB}}
\acro{SN}[SN]{supernova}
\acroplural{SN}[SN\acrolowercase{e}]{supernova}
\acro{SNIa}[\acs{SN}\,I\acrolowercase{a}]{Type~Ia \ac{SN}}
\acroplural{SNIa}[\acsp{SN}\,I\acrolowercase{a}]{Type~Ic \acp{SN}}
\acro{SNIcBL}[\acs{SN}\,I\acrolowercase{c}\nobreakdashes-BL]{broad\nobreakdashes-line Type~Ic \ac{SN}}
\acroplural{SNIcBL}[\acsp{SN}\,I\acrolowercase{c}\nobreakdashes-BL]{broad\nobreakdashes-line Type~Ic \acp{SN}}
\acro{SVD}[SVD]{singular value decomposition}
\acro{TAROT}[TAROT]{T\'{e}lescopes \`{a} Action Rapide pour les Objets Transitoires}
\acro{TDOA}[TDOA]{time delay on arrival}
\acroplural{TDOA}[TDOA\acrolowercase{s}]{time delays on arrival}
\acro{TD}[TD]{time domain}
\acro{TOA}[TOA]{time of arrival}
\acroplural{TOA}[TOA\acrolowercase{s}]{times of arrival}
\acro{TOO}[TOO]{target\nobreakdashes-of\nobreakdashes-opportunity}
\acroplural{TOO}[TOO\acrolowercase{s}]{targets of opportunity}
\acro{UFFO}[UFFO]{Ultra Fast Flash Observatory}
\acro{UHE}[UHE]{ultra high energy}
\acro{UVOT}[UVOT]{UV/Optical Telescope\acroextra{ (instrument on \emph{Swift})}}
\acro{VHE}[VHE]{very high energy}
\acro{VISTA}[VISTA@ESO]{Visible and Infrared Survey Telescope}
\acro{VLA}[VLA]{Karl G. Jansky Very Large Array}
\acro{VLT}[VLT]{Very Large Telescope}
\acro{VST}[VST@ESO]{\acs{VLT} Survey Telescope}
\acro{WAM}[WAM]{Wide\nobreakdashes-band All\nobreakdashes-sky Monitor\acroextra{ (instrument on \emph{Suzaku})}}
\acro{WCS}[WCS]{World Coordinate System}
\acro{WSS}[w.s.s.]{wide\nobreakdashes-sense stationary}
\acro{XRF}[XRF]{X\nobreakdashes-ray flash}
\acroplural{XRF}[XRF\acrolowercase{s}]{X\nobreakdashes-ray flashes}
\acro{XRT}[XRT]{X\nobreakdashes-ray Telescope\acroextra{ (instrument on \emph{Swift})}}
\acro{ZTF}[ZTF]{Zwicky Transient Facility}
\end{acronym}

\acknowledgements

See the Supplement \citep{GW150914-EMFOLLOW-SUPPLEMENT} for a full list of acknowledgements. This is LIGO document LIGO-P1500227-v12.

\bibliographystyle{aasjournal}

\end{document}